\documentclass{Interspeech}

\usepackage{listings}

\interspeechcameraready

\title{Scalable Offline ASR for Command-Style Dictation in Courtrooms}

\author[affiliation={1}]{Kumarmanas}{Nethil}
\author[affiliation={1}]{Vaibhav}{Mishra}
\author[affiliation={1}]{Kriti}{Anandan}
\author[affiliation={1}]{Kavya}{Manohar}

\affiliation{}{Adalat AI}{India}
\email{manas@adalat.ai, vaibhav@adalat.ai, kriti@adalat.ai, kavya@adalat.ai}
\keywords{Automatic Speech Recognition, Batch processing, Multiplexing, Speech Technology}

\usepackage{comment}

\begin{document}

\maketitle

\begin{abstract}
    
We propose an open-source framework for Command-style dictation that addresses the gap between resource-intensive Online systems and high-latency Batch processing. Our approach uses Voice Activity Detection (VAD) to segment audio and transcribes these segments in parallel using Whisper models, enabling efficient multiplexing across audios. Unlike proprietary systems like SuperWhisper, this framework is also compatible with most ASR architectures, including widely used CTC-based models. Our multiplexing technique maximizes compute utilization in real-world settings, as demonstrated by its deployment in around 15\% of India's courtrooms. Evaluations on live data show consistent latency reduction as user concurrency increases, compared to sequential batch processing. The live demonstration will showcase our open-sourced implementation and allow attendees to interact with it in real-time.

\end{abstract}

\section{Introduction}

Command-style dictation represents a critical workflow in environments where users need to dictate discrete segments of text with minimal latency. In judicial settings particularly, this modality enables real-time documentation of proceedings without disrupting courtroom dynamics.

Traditional ASR approaches are poorly suited for this task: Batch systems like WhisperX \cite{bain2022whisperx} process inputs sequentially, which introduces unacceptable latency for interactive applications, while online real-time ASR demands dedicated compute per user, making it prohibitively expensive to scale. Recent work has attempted to address these limitations by optimizing Whisper for real-time transcription \cite{machacek-etal-2023-turning}, but these approaches still face challenges in multi-user scenarios. Proprietary solutions like SuperWhisper have attempted to bridge this gap, but their closed-source implementations limit accessibility and widespread adoption, particularly in resource-constrained environments.

Our framework addresses these limitations through an open-source approach that decouples Voice Activity Detection (VAD) from transcription while implementing strategic multiplexing across concurrent audio streams. By treating each speech segment as an independent unit, we enable parallel processing both within individual streams and across multiple users. This design leverages VAD for precise speech segmentation, followed by parallel batch processing with fine-tuned Whisper models, that significantly improves transcription throughput without modifying core model architectures. The approach is also compatible with ASR architectures that operate without contextual decoding dependencies, including widely used CTC-based models.

Deployed in approximately 15\% of India's courtrooms, this system demonstrates effectiveness in real-world, mission-critical environments. Evaluations show it maintains transcription quality and acceptable latency even as concurrent user load increases - validating its viability for command-style dictation at scale.

The remainder of this paper details our system architecture in Section 2, presents experimental results in Section 3, and concludes with future directions in Section 4. A live demonstration of the system will accompany this paper at Interspeech.

\begin{figure}[t]
  \centering
  \includegraphics[width=\linewidth]{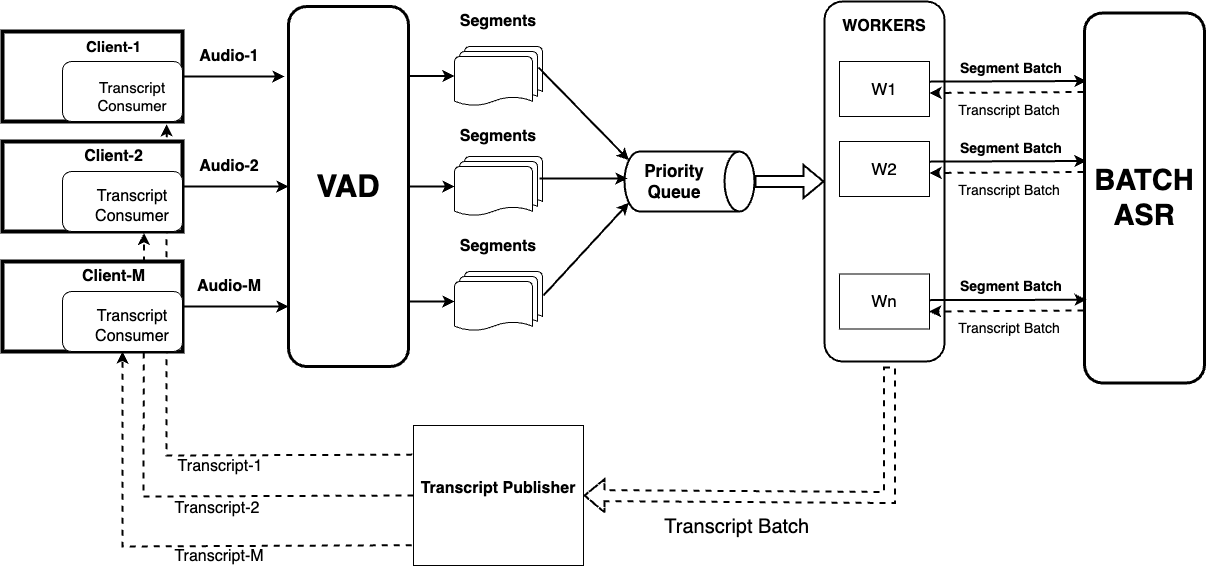}
  \caption{System architecture}
  \label{fig:architecture}
\end{figure}

\section{Approach}

We implemented a distributed architecture where VAD and ASR components operate independently. As shown in Figure 1, our system identifies discrete speech segments within each audio file, enabling parallel processing across multiple users.

\subsection{VAD Optimization}
We configure Silero VAD to produce segments between 3-30 seconds with 300ms silence padding, to align segments closely with natural speech boundaries. This calibration varies by model type: Whisper models perform optimally with longer segments and appropriate padding, while CTC-based models can process shorter segments with minimal padding. Proper VAD segmentation is crucial as it creates self-contained linguistic units that can be processed independently, directly enabling the parallel inference that our system depends on.

\subsection{Multiplexing Strategy}
Our approach utilizes a centralized priority queue to aggregate speech segments from all active users. This queue dynamically batches segments into a shared inference pipeline, transforming sequential speech recognition into a parallel processing problem. We support both dynamic batching (adjusting based on segment length and user wait times) and continuous batching (maintaining minimum batch size for consistent GPU utilization). This multiplexing significantly improves efficiency in multi-user environments by maintaining high GPU utilization even when individual users have irregular dictation patterns.

\subsection{Parallel ASR Inference}
We treat each VAD-identified segment as a self-contained unit, sidestepping the sequential dependencies in autoregressive models. This works because properly segmented commands don't require inter-segment context, allowing independent processing despite Whisper's autoregressive nature.

The following code snippet demonstrates our low-level batching of compute operations. The \texttt{prompt\_tokens} provide generation context while \texttt{no\_timestamp\_token} disables timestamp generation for faster inference.

\begin{lstlisting}[caption={Parallel inference with faster-whisper}, label={lst:paralle-inference}]
# Extract features for all segments in batch
features = torch.stack([
    torch.from_numpy(pad_or_trim(model.feature_extractor(chunk)))
    for chunk in audio_chunks
])

# Generate transcriptions in parallel
encoder_output = model.encode(features)  
generation_results = model.model.generate(
    encoder_output, 
    [prompt_tokens + no_ts_token] * len(features)
)
\end{lstlisting}

This architecture supports both Whisper and alternative ASR architectures that can process discrete segments independently, optimizing for command-style dictation where throughput is the priority.

\section{Efficiency of Multiplexing}

Our multiplexing approach transforms ASR economics in multi-user environments by dynamically batching speech segments across concurrent users. We compare this against the baseline of sequential processing, where each audio is processed completely before starting the next.

\subsection{Methodology}
We evaluated performance using 100 authentic courtroom recordings (5s-5m length) on a server with 1× NVIDIA T4 GPU. Testing covered three concurrency levels (5, 10, 20) across audio durations ranging from 15 to 120 seconds, with p90 latency as our primary metric.

\subsection{Results}
As Figure 2 shows, our multiplexing approach consistently outperforms the sequential baseline. At lower concurrency (C=5), the multiplexed system shows modest improvements of 0.5–1.0 seconds across most audio durations. At moderate concurrency (C=10), the benefits of multiplexing become more noticeable for longer audio. For 105–120s clips, sequential processing reaches a p90 latency of approximately 7.2s, whereas multiplexed processing reduces it to 6.2s—an improvement of about 14\%.

The most significant benefits appears at high concurrency (C=20). For longer audio clips (105–120s), sequential processing latency increases dramatically to approximately 13.5s, while multiplexed processing maintains a lower latency of around 10s—a 26\% improvement. This demonstrates the scalability of our approach in real-world conditions where system responsiveness under heavy load is critical.

We expect this performance gap to widen further for even longer audio segments, where the cumulative benefits of overlapping computation and reduced queuing become more pronounced.

\begin{figure}[t]
  \centering
  \includegraphics[width=\linewidth]{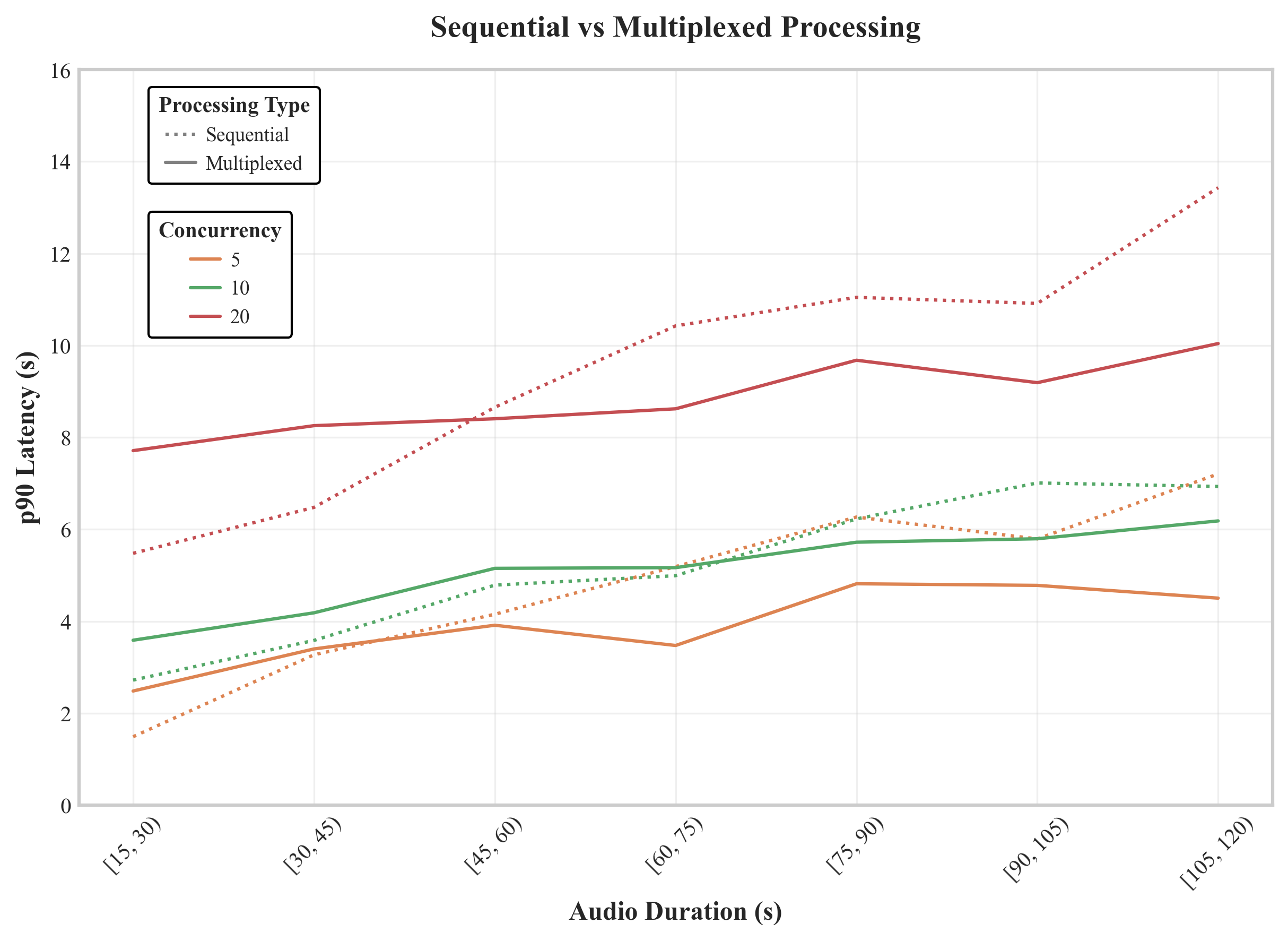}
  \caption{p90 Latency: Sequential vs Multiplexed inference across audio durations and user concurrencies}
  \label{fig:latency-vs-concurrency}
\end{figure}

\section{Conclusion}

We present an open-source ASR framework tailored for command-style dictation, combining VAD-based segmentation with parallel, multiplexed inference to achieve low-latency transcription at scale. By decoupling speech detection from decoding, the system avoids the trade-offs of traditional batch and online ASR approaches, enabling efficient multi-user performance without per-user GPU overhead.

Future work will focus on advancing the system's core architecture, including smarter batching strategies in the multiplexing queue, improved VAD-postprocessing to better align segment boundaries, and dynamic resource allocation for large-scale deployments with bursty user patterns.

Our live demonstration at Interspeech will feature a web-browser frontend alongside a server dashboard showing connected users and perceived RTF - allowing attendees to interact with the system in real time and observe how it maintains performance under concurrent load.

\bibliographystyle{IEEEtran}
\bibliography{mybib}

\end{document}